\documentclass[12pt]{article}

\pdfoutput=1

\usepackage{a4wide}
\usepackage{amssymb}
\usepackage{amsmath}
\usepackage{graphicx}
\usepackage{mathdots}
\usepackage{slashed}
\usepackage{verbatim}
\usepackage{xcolor}
\usepackage{cancel}

\usepackage{cite}

\usepackage{hyperref}

\usepackage{IEEEtrantools}

\usepackage{amsfonts}

\usepackage{epsfig}

\newcommand{\be}{\begin{equation}}
\newcommand{\ee}{\end{equation}}
\newcommand{\bea}{\begin{eqnarray}}
\newcommand{\eea}{\end{eqnarray}}
\newcommand{\ba}{\begin{eqnarray}}
\newcommand{\ea}{\end{eqnarray}}

\newcommand{\WW}{{\cal W} }

\def\XXint#1#2#3{{\setbox0=\hbox{$#1{#2#3}{\int}$}
     \vcenter{\hbox{$#2#3$}}\kern-.52\wd0}}

\begin{document}

\setcounter{table}{0}


\mbox{}
\vspace{0truecm}
\linespread{1.1}

\vspace{0.5truecm}

\centerline{\Large \bf  New exact solutions in multi-scalar field cosmology} 


\vspace{1.3truecm}

\centerline{
    {\large \bf Jorge G. Russo} }

\vspace{0.8cm}

\noindent  
\centerline {\it Instituci\'o Catalana de Recerca i Estudis Avan\c{c}ats (ICREA), }
\centerline{\it Pg. Lluis Companys, 23, 08010 Barcelona, Spain.}

\medskip
\noindent 
\centerline{\it  Departament de F\' \i sica Cu\' antica i Astrof\'\i sica and Institut de Ci\`encies del Cosmos,}
\centerline{\it Universitat de Barcelona, Mart\'i Franqu\`es, 1, 08028 Barcelona, Spain. }

\medskip

\centerline{  {\it E-Mail:}  {\texttt jorge.russo@icrea.cat} }

\vspace{1.2cm}

\centerline{\bf ABSTRACT}
\medskip

We use the method of the superpotential  to derive exact solutions describing inflationary cosmologies in multi-field models.
An example that describes a solution that interpolates between two de Sitter universes is described in detail.
New analytical solutions for axion-dilaton cosmologies are also presented.

\noindent

\newpage

\def\sech{ {\rm sech}}
\def\p{\partial}
\def\pa{\partial}
\def\ov{\over }
\def\a{\alpha }
\def\g{\gamma}
\def\s{\sigma }
\def\td{\tilde }
\def\vp{\varphi}
\def\strokedint{\int}
\def \ha {{1 \over 2}}

\def\KK{{\cal K}}




\textwidth = 460pt
\hoffset=0pt

\tableofcontents


\section{Introduction}

Inflationary cosmologies driven by the dynamics of multiple scalar fields have recently received renewed attention \cite{Bassett:2005xm,Wands:2007bd,Gong:2016qmq,Christodoulidis:2019mkj} .
A scenario with many scalar fields is natural in the context of string theory compactifications,
which typically lead to a low-energy effective field theory with multiple fields that could, in principle, participate in the dynamics of the early universe cosmology.
Multi-field cosmology may also incorporate a number of interesting effects, which have been discussed in different contexts, such as {\it e.g.}  hybrid inflation \cite{Linde:1993cn}
or rapid-turn inflation \cite{Bjorkmo:2019fls,Aragam:2019omo,Aragam:2020uqi,Kolb:2022eyn}. Cosmological models with scalar fields are also of interest because they offer a variety of generalizations of standard cosmological models while maintaining isotropy
 and homogeneity.

In general, the equations that govern the dynamics are highly non-linear and one  usually has to resort to numerical methods.
Clearly, it is of  interest to also have  realistic models
of inflation described in terms closed formulas.
There have been many relevant studies in the literature describing analytic solutions 
or studying integrable systems
(see {\it e.g.} \cite{Salopek:1990jq,Muslimov:1990be,Kinney:1997ne,Chervon:1997yz,Russo:2004ym,Arefeva:2005mka,Vernov:2006dm,Arefeva:2009tkq,Fre:2013vza,Paliathanasis:2014yfa,Chervon:2017kgn,Christodoulidis:2018msl,Dimakis:2019qfs,Chervon:2019nwq} for a very incomplete reference list).

The purpose of this note is to point out some interesting solutions that have apparently escaped the attention of previous studies. We will exploit the familiar method of the superpotential \cite{Salopek:1990jq,Muslimov:1990be}. This can be viewed as the  Hamilton's characteristic function  of Hamilton-Jacobi theory \cite{Salopek:1990jq,Muslimov:1990be,Kinney:1997ne,Skenderis:2006rr}.
Assuming isotropy and homogeneity, the Lagrangian for gravity coupled to scalar fields takes the general form
\be
L =\frac12 G_{ij} \partial_t \phi^i \partial_t \phi^j-V(\phi^i)\ .
\ee
The superpotential $\WW $, when this exists, is defined by the equation
\be
\label{potiuno}
V=-\frac12 G^{ij} \frac{\partial \WW }{\partial \phi^i}\frac{\partial \WW }{\partial \phi^j}\ .
\ee
This ensures that the constraint is satisfied and that the solutions of the first order system
\be
\label{ecuuno}
\partial_t \phi^i = G^{ij} \frac{\partial \WW }{\partial \phi^j}\ ,
\ee
solve the second-order equations.
For a given superpotential, the first-order system \eqref{ecuuno} describes a particular solution in the more general space of solutions to the second-order equations. 

\medskip

\section{Multi-field inflationary cosmology}

Consider the following Lagrangian describing
Einstein gravity coupled to $n$ scalar fields $\phi^i$
with self-interactions,
\be
\label{lagrim}
L=\frac12 \sqrt{-{\rm det}g} \left(2R- g^{\mu\nu} G_{ij}(\phi) \partial_\mu \phi^i   \partial_\nu \phi^j-2V( \phi) \right)\ .
\ee
Here we adopt units where $\kappa^2=8\pi G=1/2$.
We shall study flat Friedmann-Lema\^ itre-Robertson-Walker (FLRW) cosmologies of the form
\be
\label{flrwan}
ds^2= - e^{2\alpha\varphi } f^2 d\tau^2 +e^{2\beta \varphi} (dx_1^2+dx_2^2+dx_3^2) ,
\ee
with $\varphi=\varphi(\tau)$, $f=f(\tau)$, $\phi^i = \phi^i (\tau)$, and
\be
\alpha = \frac{\sqrt{3}}{{2}}  
\ ,\qquad \beta =\frac{1}{2\sqrt{3}}\ .
\ee
The choice $f=e^{-\alpha \varphi}$ corresponds to the standard cosmological time.

The Einstein equations and the scalar field equations of the original Lagrangian then
reduce to the Euler-Lagrange equations of the effective Lagrangian
\be
\label{logro}
L_{\rm eff}=\frac{1}{2f} \left(- \dot \varphi^2+ G_{ij}(\phi)  {\dot \phi^i}  { \dot \phi^j} \right)  -f e^{2\alpha\varphi} V( \phi)\ .
\ee
Even for very simple potentials, solving the  system of coupled second-order differential equations is very complicated.
Here we will be using a simple construction
where equations reduce to more tractable first-order differential equations.
In the standard method of the superpotential, one starts with a theory with  potential $V$ and searches for a function ${\cal W}$ satisfying \eqref{potiuno}.
Here the idea is to reverse the logic and to identify a class of  potentials that can be derived from a predetermined superpotential.
We  define the following superpotential:
\be
\label{ansatW}
\WW = e^{\alpha \varphi} W( \phi )\ .
\ee
For a judiciously chosen $W$, we then  consider a model with the potential 
\be
\label{potro}
2e^{2\alpha\varphi}\, V(\phi)\equiv (\partial_\varphi \WW)^2 -G^{ij} \partial_i \WW \partial_j \WW =e^{2\alpha \varphi} \left(\alpha^2W^2-G^{ij} 
\partial_i W\partial_j W\right)\ ,
\ee
where we adopted the shorthand notation
$\partial_i \equiv\partial_{\phi_i}$.

The superpotential needs not to be of the form \eqref{ansatW}. In fact, in section 5 we will
study a model where the dependence on $\varphi $ and $\phi^i$ does not factorize.

Let us now consider  a model with $G_{ij}=\delta_{ij}$.
The $f$ equation gives rise to the constraint:
\be
\label{contar}
\dot \varphi^2-\sum_i \dot \phi_i^2  = 2f^2 e^{2\alpha\varphi} V\ .
\ee
We shall choose the standard cosmological time $t$, corresponding to the choice  $f=e^{-\alpha\varphi}$.
The constraint takes the form
\be
\label{contar2}
\dot \varphi^2-\sum_i \dot \phi_i^2  = 2 V\ .
\ee
Here, and in what follows, dot represents derivative with respect to $t$.
The second-order equations are
\be
\ddot \varphi +\alpha \dot\varphi^2-2\alpha V(\phi)=0\ 
,\qquad
\ddot \phi_i +\alpha\dot \varphi \dot\phi_i + \partial_i V(\phi)=0\ ,
\ee
or, in terms of Hubble parameter $H=\beta \dot\varphi =\dot\varphi/(2\sqrt{3})$,
\be
\dot H +3 \, H^2-\frac12\, V(\phi)=0\ 
,\qquad
\ddot \phi_i +3H \dot\phi_i +\partial_i V(\phi)=0\ ,\quad 6H^2=\frac12 \sum_i \dot \phi_i^2+V\ .
\ee

 The potential is then defined by ($\alpha^2=3/4$)
\be
 2V = \alpha^2 W^2 - (\partial_i W)^2\ .
\label{elpoti}
\ee
Using \eqref{ecuuno}, we  find the following
 first-order system; 
\be \label{ffii}
\dot \varphi = -\epsilon \alpha W\ , \qquad \dot \phi_i = \epsilon \partial_{i} W\ .
\ee
The two possible choices of  $\epsilon= \pm 1$ correspond to the two possible choices of directions of time.
We will adopt the convention where the choice of the sign $\epsilon= \pm 1$ will be the one that leads to expansion at late times.

The Hubble parameter can then be written in terms of $\phi_i$ without time derivatives, 
\be
H=\frac{\dot a}{a}=\beta\dot \varphi=-\epsilon \, \frac{W(\phi)}{4} 
\ .
\ee
Using this formula, one can anticipate the behavior
of the Hubble function by following  the evolution of the matter fields.

Similarly, the slow-expansion parameters are also functions of the values of the matter fields $\phi_i$,
\be
\label{lentop}
\epsilon_H= -\frac{\dot H}{H^2}=\frac{4(\partial_{i} W)^2}{W^2} \ ,\qquad \eta_H =-\frac{\ddot H}{2H\dot H }= 4 \frac{\partial_i W\partial_j W \partial_{i}\partial_j W}{W(\partial_i W)^2}
\ee
The slow-roll condition $\epsilon_H\ll 1 $ 
requires $|\partial_i W|\ll |W|$, which implies $V\approx \frac38 W^2$. Using this, one can   relate $\epsilon_H,\ \eta_H$ to the standard slow-roll parameters $\epsilon_V, \ \eta_V$ defined in terms of the potential. 

 The number of e-folds occurred during the inflation period is therefore
\be
N=\int_{t_0}^{t_f} dt\,  H(t) =-\frac14 \epsilon \int_{t_0}^{t_f} dt \, W(\phi_i(t))=\frac{1}{2\sqrt{3}}\, \big(\varphi(t_f)-\varphi (t_0) \big)\ .
\ee

Let us now consider the equation of state $p=w\rho$, where $p$ is the matter pressure and $\rho $ is the energy density,   obtained from
$T_{00}=-g_{00}\rho\ ,\ \ T_{ij} = g_{ij} p\ $.
In the present case,
\be
\rho=\frac{1}{2}\left(  \sum_i\dot \phi_i^2+2V\right)\ ,\qquad
p=\frac{1}{2}\left(  \sum_i\dot \phi_i^2-2V\right)\ .
\ee
Using the constraint \eqref{contar2},
the density and pressure reduce to
\be
\rho=\frac{1}{2} \dot\varphi^2\ ,\qquad
p=\frac{1}{2}\left(  \dot\varphi^2 - 4V\right)\ .
\ee
Hence 
\be
\label{wwvv}
w=1-\frac{4V}{\alpha^2 W^2}=-1+\frac{8(\partial_i W)^2}{3W^2}=-1 +\frac23 \epsilon_H\ .
\ee
This formula shows the circumstances where  a phase with $w \approx -1$ can arise.
The general condition is $|\partial_{i} W|\ll |W| $. 
One case where this 
condition is satisfied is when the trajectory $\phi_i=\phi_i(t)$ passes through the neighborhood of a fixed point ({\it i.e.} a point where all $\dot \phi_i$ are equal to zero). In this case the kinetic energy is small as compared with the potential and the FLRW spacetime approximates de Sitter space.

However, the condition $|\partial_i W|\ll |W| $ may also be satisfied away from fixed points, at large values of the $\phi_i$'s, {\it i.e.} for trajectories going to infinity. For example, if at large values of $\phi's$ the superpotential has a power-like behavior $W\approx c\phi^n$, then $w\approx -1+ O(1/\phi^2)$. In these cases the geometry is obviously very different from de Sitter. Both kinetic and potential energies go to infinity with $\frac12|\dot\phi_i|^2\ll V$, leading to $w\approx -1$, despite the fact the acceleration rate is not constant.
Examples  will be shown below.

\section{The quartic potential}

\subsection{Single-scalar field}

A model with a single scalar field governed by
a quartic potential can be obtained with the choice
\be
W= \mu+ g  \phi^2\ .
\ee
Using \eqref{elpoti}, this gives rise to the potential
\be
V= \frac38 \mu^2+\frac14 g (3\mu-8g)\phi^2+\frac38\, g^2 \phi^4 \ .
\ee
For $g (3\mu-8g)<0$, this is a Higgs-like potential with a local maximum at $\phi=0$ and minima at $\phi=\pm \sqrt{\frac83 -\frac{\mu}{g}}$.

The cosmological evolution of a single scalar field with quartic potential  has been extensively studied in the literature and there is probably nothing new to add
to the physics it describes.
However, it is instructive to find out the exact cosmological solution that arises through the present formalism  and to see how it reproduces the expected picture.

The first order equations \eqref{ffii} (with $\epsilon=-1$) take the form
\be
\dot \phi =-2g\phi\ ,\qquad \dot\varphi =\frac12 \sqrt{3}(\mu+g\phi^2) \ .
\ee
They are easily solved by direct integration. We obtain
\be
\phi= \phi_0\, e^{-2gt}\ ,\qquad \varphi =\varphi_0+\frac12 \sqrt{3}\, \mu t -\frac18 \sqrt{3}\phi_0^2\, e^{-4g t}\ .
\ee
One can easily check that the second-order equations are satisfied.
This is an interesting solution that describes a scalar field rolling down the potential from $\phi=\infty$ at $t=-\infty$, passing through the minimum of the potential and ending at the local maximum $\phi=0$ at $t=\infty$, where the universe is de Sitter. The scalar field loses energy due to Hubble friction. 

The scale factor is $a= e^{\beta \varphi}$.
The universe expands if $\dot a = \beta e^{\beta \varphi}\dot \varphi >0$. Since $\dot \varphi =\alpha W$, periods of expansion occurs when the trajectories go through a region where $W>0$.

The Hubble parameter is given by
\be
H=\frac{\dot a}{a}=\beta\dot \varphi=\frac{\mu}{4} +\frac{1}{4}\, g\phi_0^2\, e^{-4gt}\ .
\ee
The universe is expanding all the way provided $\mu>0$ and $g>0$.
We may assume the time direction such that the universe expands at late times, so we choose $\mu>0$. Then, if $g<0$, the universe undergoes contraction followed
by a final period of expansion.

Computing the slow-roll parameters $\epsilon_H$, $\eta_H$ one finds that they are $\ll 1$ provided $g\ll\mu $.
In this regime, the potential only has an absolute minima at $\phi=0$, and the solution describes the field $\phi$ rolling down from infinity to $\phi=0$.
The number of e-folds during the de Sitter period, $t>O(1/g)$, diverges, since the  de Sitter expansion continues until $t=\infty$.

If $g (3\mu-8g)<0$, then $\phi=0$ is a local maximum. In this case $\phi$ rolls down from $\phi=\infty$, passes through a minimum of the potential and  reaches the ``top of the hill" at $\phi=0$. 
If $g (3\mu-8g)\geq 0$, then $\phi=0$ is the absolute minimum of the potential, and it is also the endpoint of the cosmological evolution at $t=\infty$.

The acceleration rate is 
\be
\frac{\ddot a}{a}=\beta (\ddot\varphi +\beta\dot\varphi^2) =\frac{1}{16} \left(\mu^2 +2g\phi_0^2 (\mu -8g)\, e^{-4gt} +g^2\phi_0^4 e^{-8gt}\right) \ .
\ee
The universe starts with an accelerated expansion (if $g>0$) with infinite acceleration  at $t=-\infty$, the acceleration decreases up to a minimum value
$\ddot a/a=g(\mu-4g)$ (which can be either sign) and then increases to finally reach the asymptotic value $\mu^2/16$, where the geometry
approaches de Sitter (with a cosmological constant determined by the value of the potential at $\phi=0$).

Using the explicit form of $V$ and $W$, from \eqref{wwvv} we get
\be
\label{wones}
w=-1+\frac{32g^2 \phi^2}{3(\mu+g\phi^2)^2}= -1+\frac{32g^2 \phi_0^2 e^{-4gt}}{3(\mu+g\phi_0^2 e^{-4gt})^2} \ .
\ee
Initially, when $t\to -\infty$, $w\approx -1 +\frac{32}{3\phi_0^2}\, e^{4gt}\approx -1$. At later times, $w$ increases, reaches a maximum value
$w=-1+8g/3\mu$ and finally approaches $w=-1$ at $t\to\infty$.

\smallskip

This example exhibits the two different circumstances described in section 2 where phases with $w\approx -1$ can arise: at early times, $t\to -\infty$,
both kinetic and potential energies are large but the potential energy dominates, producing $w\approx -1$. The scale factor is very small, with $a\approx \exp(-\frac{\phi_0^2}{16}e^{-4g t})$ and the geometry is obviously not de Sitter. At late times, the kinetic energy goes to zero and the scalar field sits  at $\phi=0$, and the cosmology is equivalent to a de Sitter cosmology with cosmological constant equal to
$V(0)=\frac38\mu^2$.

\subsection{Two scalar fields}

The generalization of the previous model to two scalar fields is straightforward.
We now consider
\be
\label{wdosca}
W=\mu+ g_1 \phi_1^2+ g_2\phi_2^2 \ .
\ee
This corresponds to the potential
\begin{equation}
V= \frac38 \mu^2+\frac14 \left(g_1 (3\mu-8g_1)\phi_1^2+g_1 (3\mu-8g_2)\phi_2^2\right)
+\frac38\, \left(g_1 \phi_1^2+g_2 \phi_2^2\right)^2   \ . 
\end{equation}
If both $g_{1,2}\leq 3\mu/8$, then there is only one absolute minimum at  
$\phi_1=\phi_2=0$. When either $g_{1}> 3\mu/8$ or $g_{2}> 3\mu/8$, non trivial minima appear at $(\phi_1,\phi_2)$ equal to
$$
(\pm \sqrt{\frac83 -\frac{\mu}{g_1}},0) \quad ,\qquad (0,\pm \sqrt{\frac83 -\frac{\mu}{g_2}})\ .
$$
Fixed points for trajectories in the space $(\phi_1,\phi_2)$ are determined by the equations 
$$
\dot\phi_1=-\partial_{\phi_1}W= -2g_1\phi_1\ ,
\qquad 
\dot\phi_2=-\partial_{\phi_2}W=-2g_2\phi_2\ .
$$
This gives the origin as the only  fixed point. The fixed point is attractive when both $g_1>0$ and $g_2>0$, otherwise it has repulsive directions.

The exact solution obtained by integrating the first order equations is
\be
\phi_1= \phi_{10}\, e^{-2g_1t}\ ,\qquad \phi_2= \phi_{20}\, e^{-2g_2t}\ ,
\ee
\be
\varphi =\varphi_0+\frac12 \sqrt{3}\, \mu t -\frac18 \sqrt{3}\phi_{10}^2\, e^{-4g_1 t}-\frac18 \sqrt{3}\phi_{20}^2\, e^{-4g_2 t}\ .
\ee
The analysis of the resulting cosmology is similar to the one-scalar field case.
The Hubble parameter is
\be
H= \frac{\mu}{4} +\frac{1}{4}\, g_1\phi_{01}^2\, e^{-4g_1t}+ \frac{1}{4}\, g_2\phi_{02}^2\, e^{-4g_2t}\ .
\ee
If $g_1>0$ and $g_2>0$, at $t\to\infty $ the trajectory in the space $(\phi_1,\phi_2)$ terminates at the fixed point $(0,0)$, where we have a de Sitter cosmology.

One new feature is the case $g_1 g_2<0$, where the origin is a repulsive fixed point and the potential has flat directions. The trajectory does
not terminate at $(0,0)$,
so there is never a de Sitter phase.
The universe begins with expansion and ends with contraction
(or the other way around, if one uses the time-reversed solution).

The expansion acceleration rate is given by the formula 
\be
\frac{\ddot a}{a}=\beta (\ddot\varphi +\beta\dot\varphi^2) =\frac{1}{16}  \left(\mu+ g_1\phi_{10}^2 e^{-4g_1t}+ g_2\phi_{20}^2 e^{-4g_2t}\right)^2
-g_1^2\phi_{10}^2 e^{-4g_1t}- g_2^2\phi_{20}^2 e^{-4g_2t}\ .
\ee
which becomes constant at $t\to\infty $ if $g_1>0$ and $g_2>0$.
Now
\be
w= -1+\frac{32(g_1^2 \phi_{10}^2 e^{-4g_1t}+g_2^2 \phi_{20}^2 e^{-4g_2t})}{3(\mu+g_1 \phi_{10}^2 e^{-4g_1t}+g_2 \phi_{20}^2 e^{-4g_2t})^2} \ .
\ee
$w$ has a similar behavior as in the one scalar case \eqref{wones},  barring the case $g_1g_2<0$, where the geometry does not approximate de Sitter
at asymptotic times.

\medskip

Finally, let us comment on models with more general polynomial interactions, where the superpotential is of the form
\be
W=\mu+c_1\phi_1+c_2\phi_2+ g_1 \phi_1^2+ g_2\phi_2^2 +\gamma \phi_1\phi_2\ .
\ee
This leads to a potential $V$ with linear, quadratic, cubic and quartic  interactions. However, for generic $g_1$ and $g_2$, the $\gamma $ term can be removed by an orthogonal rotation in the space $(\phi_1,\phi_2)$. Then the linear terms with coefficients $c_1$ and $c_2$ can be removed by
 shifting $\phi_{1}$ and $\phi_2$ by an appropriate constant,
 so the model is indeed  equivalent to the one considered above with superpotential \eqref{wdosca}.\footnote{In the special case
$
W=\mu +m\phi_2 +g\phi_1^2 $ the linear term cannot be removed. 
However, this gives rise to a potential that is unbounded from below and will not be considered here.}

\subsection{Multi-scalar fields}

The generalization to the case of $N$ scalar fields is straightforward. One considers the superpotential
\be
W=\mu+ \sum_{i=1}^N g_i \phi_i^2 \ .
\ee
The first order equations are solved by
\begin{equation}
    \phi_i= \phi_{i0}e^{-2g_i t} \ ,
\end{equation}
\be
\varphi =\varphi_0+\frac12 \sqrt{3}\, \mu t -\frac18 \sqrt{3}
\sum_{i=1}^N \phi_{i0}^2\, e^{-4g_i t}\ .
\ee
with the Hubble constant given by
\be
H= \frac{\mu}{4} +\frac{1}{4}\, \sum_{i=1}^N 
g_i\phi_{i0}^2\, e^{-4g_it}\ .
\ee
The origin is an attractive fixed point only when all $g_i>0$, in which
case the geometry approaches de Sitter at late times.


\section{Non-Linear first-order systems}

Using the method of the superpotential one can also find exact solutions that are not of exponential form.

A conspicuous example was found in \cite{Russo:2022gdj},
a three-scalar model inspired by  D-branes  that leads to a chaotic cosmology described by none other than the (generalized) Lorenz strange attractor, which in the 60's gave birth to the well-known ``butterfly effect".
The potential can be derived from the superpotential
\be
\label{superw}
W= \frac12 \left(a \phi_1^2+b \phi_2^2-c \phi_3^2\right) + \phi_1 \phi_2 \phi_3\ .
\ee
and sigma-model metric $G_{ij}={\rm diag}(1,1,-1)$ for the kinetic terms for the scalar fields $(\phi_1,\phi_2,\phi_3)$.
This gives  the first-order equations,
\be
\dot \phi_1 = a\phi_1 +\phi_2 \phi_3\ , \qquad 
\dot \phi_2 = b \phi_2 +\phi_1 \phi_3\ , \qquad
\dot \phi_3 = c\phi_3 -\phi_1 \phi_2\ .
\ee
The trajectories describe a strange attractor. The model provides an example of deterministic chaos
in General Relativity. 

A crucial feature of the model is the existence of fixed points that have both repulsive and attractive directions, and the nonexistence of fixed points that are attractive in all directions. These are  essential ingredients for trajectories to describe a strange attractor. The absence of an attractive fixed point also implies that there are no trajectories that can approach a de Sitter universe at late times. The trajectories are quasi-periodic and approach de Sitter behavior only when they pass through the vicinity of fixed points, where $(\partial_i W)^2\ll W^2$ and there is dominance of 
potential energy ({\it cf.} \eqref{wwvv}). Quasi-periodic orbits go around the different fixed points interpolating in a non-monotonic way between approximate de Sitter cosmologies \cite{Russo:2022gdj}.

\smallskip

It is important to note that strange attractors cannot occur if the kinetic term is positive definite. The reason is that the  equation $\dot H = - \frac14 G_{ij} \dot \phi^i \dot\phi^j$ implies $\dot H<0$ at all times if the kinetic term is positive definite, and this makes it impossible for the system to have the quasi-periodic trajectories that characterize strange attractors. Therefore they can only occur if there is at least one ghost in the
low-energy effective Lagrangian, like in the above example.\footnote{Cosmological models inspired in D-brane dynamics with similar superpotential have also been investigated in \cite{Ashoorioon:2009wa,Ashoorioon:2009sr}, but only with positive definite kinetic terms. Other discussions of superpotentials for cosmological models with ghost scalar fields can be found in \cite{Arefeva:2005mka,Vernov:2006dm}}.

\medskip

\medskip

In this paper we  only consider examples with positive definite kinetic terms and bounded potentials. 
In particular, bounded potentials can be easily obtained by
choosing a separable superpotential of the form
\be
W=\sum_i F_i(\phi_i)\ ,
\ee
with   polynomial $F_i$'s. In addition, in these cases the first-order equations can be solved by direct integration.

One can also consider 
 superpotentials that are not separable. For example, if
\be
W=h(\phi_1)\phi_2+b \phi_2^3\ ,
\ee
where $h(\phi_1)$ is an arbitrary function, the differential equations can be  explicitly solved in closed form,
although a polynomial $h(\phi_1 )$ now gives unbounded directions.
Another example of non-separable superpotential is discussed in section 5.

\subsection{Interpolating cosmology between two de Sitter phases}

Let us begin with a single scalar field. We take the superpotential 

\be
\label{superds}
W=\mu + g\phi ^2 -\lambda\phi^4\ .
\ee
{}Using \eqref{elpoti}, one obtains the potential
\be
\label{pott}
V=\frac{3}{8} \left(g \phi^2+\mu -\lambda  \phi^4\right)^2-2\phi^2 \left(g -2 \lambda  \phi^2\right)^2\ .
\ee
The first-order equation \eqref{ffii} is now
\be
\label{awe}
\dot \phi=-2g \phi+4\lambda \phi^3\ .
\ee
A very interesting cosmology arises for $\mu>0$, $g>0$ and $\lambda>0$.
There is an attractive fixed point  at $\phi=0$ and repulsive fixed points at $\phi_{\pm}\equiv \pm \sqrt{g/(2\lambda)}$, where 
the potential has   local maxima.
At the origin, one has $V'(0)=0$, $V''(0)=\frac{1}{2} g (3 \mu -8 g)$, hence $\phi=0$ is a maximum for $\mu<8g/3$ and a minimum for
$\mu\geq 8g/3$, with
\be
\label{vcero}
V(0)= \frac{3 \mu ^2}{8}\ , \qquad V(\phi_\pm ) =\frac{3 \left(g^2+4 \lambda  \mu \right)^2}{128 \lambda ^2}\ .
\ee
In addition, the potential has  absolute minima at larger $\phi=\pm \phi_1$, with 
$$
\phi_1^2=8+
\frac{g}{2 \lambda }+\frac{1}{2\sqrt{3} \lambda }
\sqrt{3 g^2+64 g \lambda +12 \lambda  (64 \lambda +\mu )}\ ,
$$
where $V(\pm \phi_1)$ is negative definite (therefore the constant solutions $\phi=\pm \phi_1$ represent  anti de Sitter vacua of the theory).

\medskip

\begin{figure}[h!]
 \centering
 \begin{tabular}{cc}
\includegraphics[width=0.4\textwidth]{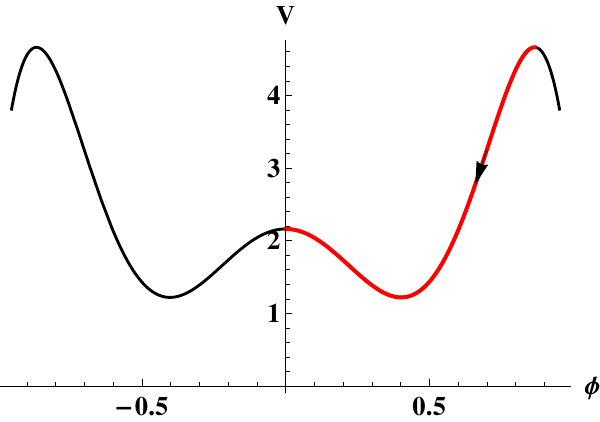}
 &
 \qquad \includegraphics[width=0.4\textwidth]{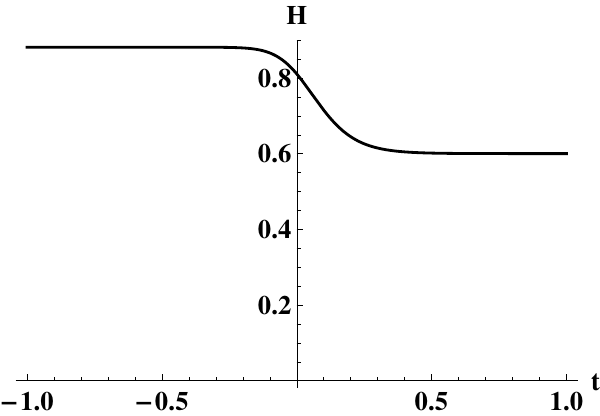}
 \\ (a)&(b)
 \end{tabular}
 \caption
 {(a) When $\mu<8g/3$, the solution \eqref{solids} describes a scalar field travelling between a maximum at 
 $\phi=\sqrt{g/(2\lambda)}$ and a local maximum at $\phi=0$ (here $g=3$, $\lambda=2$, $\mu=0.3\times 8g/3$). If $\mu\geq 8g/3$, $\phi=0$ becomes a local minimum. There are two symmetric absolute minima at
  larger $|\phi|$ not shown in this plot. For large $|\phi|$, $V$ increases as $\lambda^2\phi^8$.
  (b) The evolution of the Hubble constant for the same values of parameters.
 }
 \label{hubblep}
 \end{figure}

The solution to \eqref{awe} is
\be
\label{solids}
\phi=\sqrt{\frac{g}{2\lambda}}\, \frac{1}{\sqrt{e^{4g t} +1}} \ ,
\ee
\be
\label{solidm}
ds^2=-dt^2+a^2(dx_1^2+dx_2^2+dx_3^2)\ ,\qquad a=e^{\frac{\mu 
   t}{4}}\left(1+e^{-4 g t}\right)^{-\frac{g}{64 \lambda }} e^{-\frac{g}{64 \lambda  \left(e^{4 g t}+1\right)}} \ .
\ee
Here an integration constant has been removed by a shift of $t$.
As $t$ goes from $-\infty$ to $\infty$, $\phi$ moves from the local maximum at $\phi= \sqrt{g/(2\lambda)}$
to  $\phi=0$, which  is a maximum for $\mu<8g/3$ and a minimum for
$\mu\geq 8g/3$. There is of course  a similar solution going from $\phi= -\sqrt{g/(2\lambda)}$ to $\phi=0$, and also their time reversed versions.
On the other hand, we get
\be
H=\beta\dot\varphi =\frac14\, W =\frac{\mu}{4}+\frac{g^2 \left(2 e^{4 g t}+1\right)}{16 \lambda  \left(e^{4 g t}+1\right)^2}\ ,
\ee
which interpolates between two values
$$
H(-\infty)=\frac{\mu}{4}
 +\frac{g^2}{16\lambda}\ ,\qquad H(\infty)=\frac{\mu}{4}\ .
$$
The universe expands all the way from $t=-\infty$ to $\infty $,
corresponding to an interpolation between de Sitter cosmologies with cosmological constants given by $V(\phi_\pm )$ and $V(0)$. See fig. \ref{hubblep}.
The number of e-folds during the whole period is infinity.

\begin{figure}[h!]
 \centering
 \begin{tabular}{cc}
\includegraphics[width=0.4\textwidth]{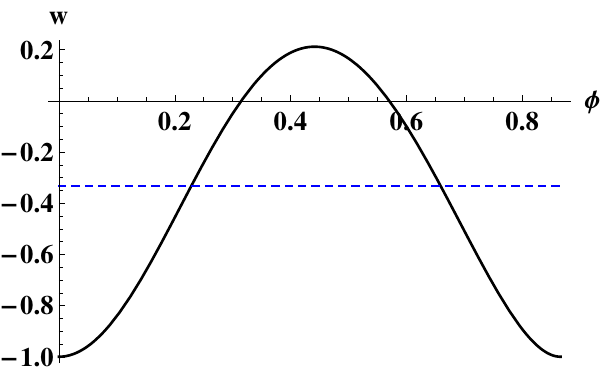}
 &
 \qquad \includegraphics[width=0.4\textwidth]{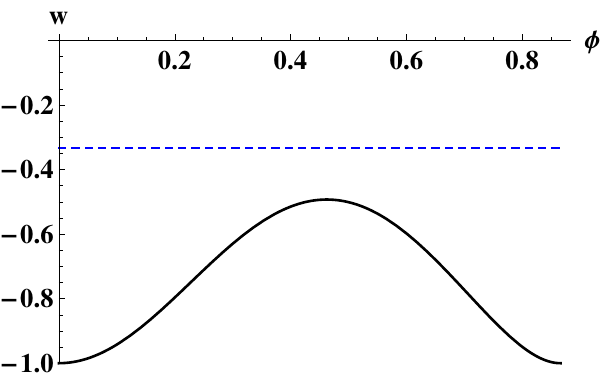}
 \\ (a)&(b)
 \end{tabular}
 \caption
 {Evolution of $w$ as the scalar field travels between the two fixed points.
 (a)  $g=3$, $\lambda=2$, $\mu=0.3\times 8g/3$. During a transient period of time, $w>-1/3$ and the expansion is decelerating.
  (b) Same $g,\lambda$, but $\mu=0.5\times 8g/3$. Now the expansion is accelerated
  all the way from $t=-\infty$ to $t=\infty$.
 }
 \label{equstate}
 \end{figure}

The expansion is accelerated but if $g$ is sufficiently large, or $\mu $ is sufficiently small, there can be a transient period of decelerating expansion. 
From \eqref{wwvv}, we find that the evolution of the equation of state parameter $w$ is
given by
\be
w=-1+\frac{32 \phi^2\left(g  -2 \lambda  \phi ^2\right)^2}{3 \left(g \phi
   ^2-\lambda  \phi ^4+\mu \right)^2}\ .
\ee
$w$ goes from $-1$ to $-1$ as $\phi$ travels between the two fixed points, passing through a maximum value (see fig. \ref{equstate}). The slow-roll parameter $\epsilon_H$ in \eqref{lentop} is $\ll 1$ at large $|t|$. On the other hand,
demanding that $\eta_H$ is small at both asymptotic $t= \pm\infty$ times require $g\ll \mu$.

The analytic solution \eqref{solids} obviously requires fine-tuning in the initial conditions, since the scalar field must have  zero kinetic energy at $t\to-\infty$ when it is sitting on the local maximum at $\phi= \sqrt{g/(2\lambda)}$.
However, it gives a hint on the cosmological evolution for more general initial conditions, using the remarkable feature  that in \eqref{solids} the evolution exactly ends at   $\phi=0$.\footnote{It is important to note that $\phi=0$ is an attractive fixed point of the first-order equation \eqref{awe}, 
but for general initial conditions it is not an attractor of the full second-order equations. That is,
the derivatives of $\phi$ generally do no vanish at $\phi=0$, so generically the scalar can pass through this point.}
Thus, if the scalar field has initially any small $|\dot\phi| $ with $\dot\phi <0$ at the moment it goes through the local maximum at $\phi= \sqrt{g/(2\lambda)}$, then, for $\mu<  8g/3$ (when $\phi=0$ is a local maximum) it will pass
through $\phi=0$ and roll down to the local minimum found at negative $\phi $ (see fig. 1a).
For $\mu\geq  8g/3$, after a few oscillations the evolution must eventually end at  $\phi=0$, which is  a local minimum in this case. Therefore the endpoint of the evolution is the same de Sitter universe, which shows that the solution is stable  under
small deformations with $\dot\phi<0$ provided $\mu\geq  8g/3$.
If, on the other hand, the scalar field initially has $\dot\phi>0$ at the time it passes through $\phi= \sqrt{g/(2\lambda)}$, then it is clear that subsequently $\phi $ will roll downhill towards the absolute minimum which is found at higher values of $\phi$. 

\medskip

When $\lambda <0$ and $g>0$, the solution is of the form
\be
\label{negativog}
\phi=\pm \sqrt{\frac{g}{2|\lambda|}}\, \frac{1}{\sqrt{e^{4g t} -1}}\ .
\ee
In this case the  trajectory starts at $t=0$ coming from   $\phi=\pm \infty$ and ends at $\phi=0$ at $t=\infty$. The geometry is singular at $t=0$ and approaches de Sitter at late times. The case $\lambda >0$ and $g<0$ corresponds to the time-reversed solution, as can be seen from \eqref{awe}.
Similarly, the case $\lambda <0$ and $g<0$ is equivalent to the time-reversed solution of the case $\lambda >0$ and $g>0$.

\subsection{The domain-wall correspondence}

Given a solution interpolating two de Sitter vacua, one can construct a `dual' solution
describing a domain wall interpolating between two anti de Sitter (adS) vacua \cite{Skenderis:2006jq}.\footnote{We thank P. Townsend for this suggestion.}
We consider the same Lagrangian \eqref{lagrim} and a domain-wall ansatz
\be
\label{adsit}
ds^2=  e^{2\alpha\varphi } f^2 dz^2 +e^{2\beta \varphi} (-dx_0^2+dx_1^2+dx_2^2) \ .
\ee
Compared with the cosmological ansatz, 
the effective Lagrangian now has an overall minus sign in the kinetic terms,
\be
\label{logroz}
L_{\rm eff}=\frac{1}{2f} \left( (\partial_z \varphi)^2-    \partial_z \phi   \partial_z \phi \right)  -f e^{2\alpha\varphi} V( \phi)\ .
\ee
Choosing now the same superpotential \eqref{superds}, the formula \eqref{potiuno} gives
a potential with an overall flip of sign in $V$,
\be
\label{potz}
V=-\frac{3}{8} \left(g \phi^2+\mu -\lambda  \phi^4\right)^2+2 \phi^2\left(g -2 \lambda  \phi^2\right)^2\ .
\ee
As a result, the signs in \eqref{vcero} are also flipped and the construction now yields a solution interpolating between two adS metrics,
representing the IR and UV regimes,
\be
\label{adsIR}
ds^2_{\rm IR}=   dz^2 +e^{2 z/L_1} (-dx_0^2+dx_1^2+dx_2^2) \ ,\qquad
{\rm for}\ z\to -\infty\ ,
\ee
and
\be
\label{adsUV}
ds^2_{\rm UV}=   dz^2 +e^{2z/L_2} (-dx_0^2+dx_1^2+dx_2^2) \ ,\qquad
{\rm for}\ z\to \infty\ ,
\ee
with adS radii given by
\be
  L_1= \frac{16\lambda}{4\mu\lambda+ g^2}\ ,\qquad L_2= \frac{4}{\mu}\ .
\ee
In the holographic description, the solution describes a renormalization group flow between
a UV CFT to another IR CFT.
The complete solution for $\phi$ and $\varphi$ is as in \eqref{solids}, \eqref{solidm},
formally changing  $t\to z$.

It would  be interesting to see if this is a `supersymmetric' domain-wall solution
in the sense of ``fake" supergravity and also to explore potential holographic applications.

\subsection{Case of two scalar fields}

The previous model can be readily generalized to the case of  two scalar fields.
 The superpotential is
\be
W=\mu + g_1\phi_1^2 -\lambda_1\phi_1^4+ g_2\phi_2^2 -\lambda_2\phi_2^4\ .
\ee
This leads to a potential with $O(\phi^8)$ terms, with many interaction terms mixing $\phi_1$ and $\phi_2$, such as $\phi_1^2\phi_2^4$, etc.
However, the first-order equations are decoupled,
\be\label{traye}
\dot\phi_1=-2g_1\phi_1+ 4\lambda_1\phi_1^3 \ ,\qquad 
\dot\phi_2=-2g_2\phi_2+ 4\lambda_2\phi_2^3 \ .
\ee
The case with $g_{1,2}>0, \ \lambda_{1,2}>0$, leads to  cosmologies that are  similar to the one-field case discussed above.
There are fixed points at $(\phi_1,\phi_2)$ equal to
\be
(0,0)\ ,\quad \Big(\pm \sqrt{\frac{g_1}{2\lambda_1}},0\Big)\ ,\quad \Big(0,\pm \sqrt{\frac{g_2}{2\lambda_2}}\Big)\ ,\quad \Big(\pm \sqrt{\frac{g_1}{2\lambda_1}},\pm \sqrt{\frac{g_2}{2\lambda_2}}\Big) \ .
\label{repuls}
\ee
The only attractive fixed point is $(0,0)$. The solutions to \eqref{traye} describe eight different trajectories going from
one of the above fixed points with non-zero $\phi_1$ or $\phi_2$, to $(0,0)$, 
thus interpolating between two de Sitter universes as time flows from minus infinity to infinity.

\begin{figure}[h!]
 \centering
 \begin{tabular}{cc}
\includegraphics[width=0.4\textwidth]{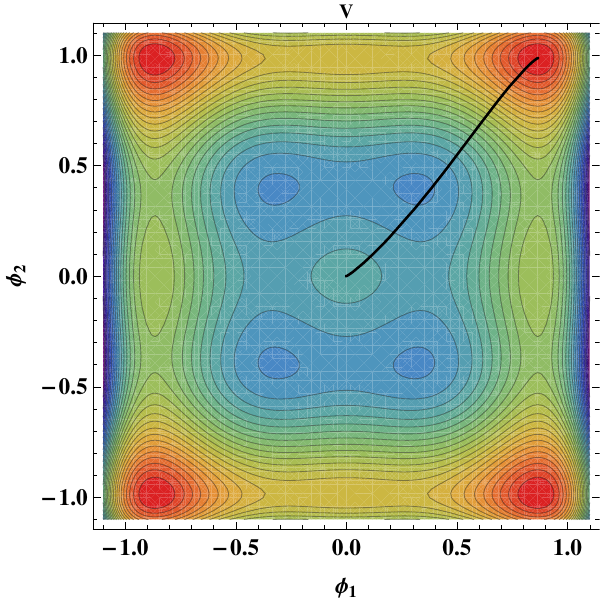}
 &
 \qquad \includegraphics[width=0.4\textwidth]{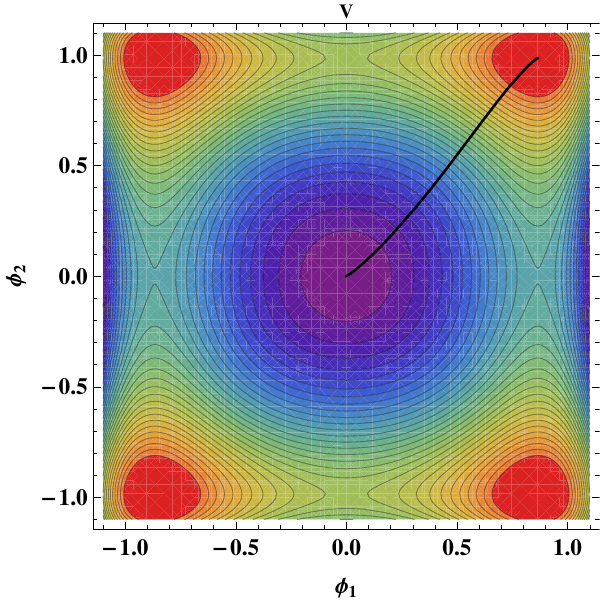}
 \\ (a)&(b)
 \end{tabular}
 \caption
 {Contour lines of the potential -- with red color indicating higher $V$, blue color indicating smaller $V$, and the trajectory (solid black line) for the solution \eqref{solidtwo}. (a) When $\mu<8g_1/3$ and $\mu<8g_2/3$, the origin is a local maximum
 and the trajectory passes through a minimum of the potential  (here $g_1=3$, $\lambda_1=2$, $g_2=3.5$, $\lambda_2=1.8$ $\mu=4$). b) When $\mu\geq 8g_1/3$ and $\mu\geq 8g_2/3$, the origin becomes a local minimum (same values for $g_{1,2},\ \lambda_{1,2}$, but now $\mu=16$). 
 }
 \label{pot2d}
 \end{figure}

The potential depends on five parameters and the structure of minima and maxima changes as
these parameters are varied.
The four fixed points $\Big(\pm \sqrt{\frac{g_1}{2\lambda_1}},\pm \sqrt{\frac{g_2}{2\lambda_2}}\Big)$
are local maxima, while the other four fixed points $\Big(\pm \sqrt{\frac{g_1}{2\lambda_1}},0\Big)\ ,\quad \Big(0,\pm \sqrt{\frac{g_2}{2\lambda_2}}\Big)$ can be local maxima or saddle-points, according to the value of the parameters. The contour lines of the potential can be visualized in figs. \ref{pot2d}a,b.
There are in addition four minima at
  larger $|\phi_{1,2}|$ not shown in these plots. The potential is bounded from below; for large $|\phi_{1,2}|$, $V$ increases as $\frac38(\lambda_1^2\phi_1^8+\lambda_2^2\phi_2^8)$.
  
Let us consider an example.
The solution to \eqref{awe} interpolating between the fixed point at $\Big( \sqrt{\frac{g_1}{2\lambda_1}}, \sqrt{\frac{g_2}{2\lambda_2}}\Big)$ and the fixed point at $(0,0)$ is given by
\be
\label{solidtwo}
\phi_1=\sqrt{\frac{g_1}{2\lambda_1}}\, \frac{1}{\sqrt{e^{4g_1 t} +1}} \ ,\qquad \phi_2=\sqrt{\frac{g_2}{2\lambda_2}}\, \frac{1}{\sqrt{e^{4g_2 t} +1}}\ ,
\ee
with the scale factor given by
\be
\label{solidmore}
a=e^{\frac{\mu 
   t}{4}}\left(1+e^{-4 g_1 t}\right)^{-\frac{g_1}{64 \lambda_1 }} e^{-\frac{g_1}{64 \lambda_1  \left(e^{4 g_1 t}+1\right)}}
   \left(1+e^{-4 g_2 t}\right)^{-\frac{g_2}{64 \lambda_2 }} e^{-\frac{g_2}{64 \lambda_2  \left(e^{4 g_2 t}+1\right)}}\ .
\ee
It is worth stressing that this straightforward generalization of the one-scalar field solution
actually solves a highly complicated coupled, non-linear system of 
second-order differential equations.
The solution again represents an accelerated expanding cosmology, interpolating
between two de Sitter universes.

Small perturbations of the initial conditions can be analysed similarly as in the one-field case, 
by a close examination of the potential. 
When $\mu\geq 8g_1/3$ and $\mu \geq 8g_2/3$, the potential has the shape of fig. \ref{pot2d}b,
where the origin is a minimum. Any trajectory having initial `velocity' vector $(\dot\phi_1,\dot\phi_2)$
pointing  inwards (that is, with $\dot\phi_1<0,\ \dot\phi_2<0$) will eventually fall at the local minimum at $(0,0)$, and the cosmology will approach the same de Sitter universe. 
If both $\mu< 8g_1/3$ and $\mu < 8g_2/3$, the origin is a local maximum
and inward trajectories will eventually fall into one of the four blue valleys of fig. \ref{pot2d}a.
Similarly, if $\mu< 8g_1/3$ and $\mu \geq 8g_2/3$ or viceversa, the origin is a saddle-point so the
trajectory will also end in one of the blue valleys. Since $V>0$ at the blue valleys, the universe again
approaches de Sitter at late times, but now with a smaller cosmological constant.
On the other hand, outgoing trajectories having either $\dot\phi_1>0$ or $\dot\phi_2>0$ will eventually
fall into one of the minima located at higher $|\phi_1|,\ |\phi_2|$, where $V<0$, so for these trajectories
the cosmology approaches anti-de Sitter at late times.

\medskip

 When $\lambda_1<0$ or $\lambda_2<0$, the picture is similar to the single scalar field case \eqref{negativog}, with the trajectory coming from   $\phi_1$ or $\phi_2=\pm\infty$  and ending at the origin  $\phi_1=\phi_2=0$, at late times. Other sign choices, such
 as $g_1<0$ or $g_2<0$, can be studied in a similar way
 (taking into account that opposite sign choices describe the time-reversed cosmology).
 The existence of fixed points away from the origin
 depends on the sign of $g_1/\lambda_1$ and $g_2/\lambda_2$.
 When both are negative, the only fixed point is the origin, which is repulsive for $g_{1,2}<0$.
 
 Finally, in the case $\lambda_1=0$, $\lambda_2>0$ (or vice-versa),
 the only fixed point is again the origin.
 The solution for $\phi_1$ is the exponential solution described in section 3.1, and $\phi_2 $ is as given in \eqref{solidtwo}. 
 The cosmology then describes an expanding universe
 that approaches de Sitter at late time, when the trajectory
 meets the origin $(0,0)$.
The corresponding scale factor is 
\be
\label{solidcero}
a=e^{\frac{\mu 
   t}{4}}\ e^{- \phi_{0}^2 e^{-4g_1 t} }
   \left(1+e^{-4 g_2 t}\right)^{-\frac{g_2}{64 \lambda_2 }} e^{-\frac{g_2}{64 \lambda_2  \left(e^{4 g_2 t}+1\right)}}\ .
\ee


\section{Cosmological models inspired from  Supergravity}

Models with exponential potentials are common in Supergravity theories. Some models can be accommodated
by an ansatz of the form
\be
W=\sum_k c_k \ e^{a_{1,k} \phi_1}...e^{a_{n,k} \phi_n}\ .
\ee
This gives rise to a potential with a number of terms with exponential dependence on the fields.
It is then easy to derive exact solutions by solving the first-order equations.

There is also a large family of models arising
from massive supergravity and from flux compactification containing  axion and dilaton fields (see {\it e.g.} \cite{Cowdall:1996tw,Russo:2022pgo}).
As an example, consider the following Lagrangian
\be
\label{teorla}
L=\frac12 \sqrt{-{\rm det}g} \left(2R- \partial_\mu \phi \partial^\mu \phi-e^{\mu\phi}\partial_\mu \chi \partial^\mu \chi-2m^2e^{\lambda\phi} \right)\ .
\ee
The Lagrangian has the standard shift symmetry of the axion field, $\chi\to\chi+{\rm const.}$ Such Lagrangians and similar ones have been investigated in \cite{Dimakis:2019qfs,Paliathanasis:2020sfe,Sonner:2006yn,Russo:2022pgo,Marconnet:2022fmx}.
For a flat FLRW ansatz, the equations of motion reduce to an autonomous system, and the
 entire cosmological evolution can be described in terms of trajectories, once the fixed points have been identified \cite{Sonner:2006yn,Russo:2022pgo} 
  (see also similar approaches in \cite{Odintsov:2017tbc}).
  Examples of  particular analytic solutions for a model with $\mu $ proportional to $\lambda $ can be found in section 3b of \cite{Dimakis:2019qfs}.

We will now show that in a particular case one can also find new solutions in closed form using the method of the superpotential.
%
We consider the FLRW ansatz \eqref{flrwan}.
To begin with, it is convenient to solve 
the equation of motion for $\chi=\chi(t)$,
\be
\dot \chi =J f e^{-\mu\phi} \ ,
\ee
where $J$ is an integration constant.
The remaining equations for $\varphi$ and $ \phi$ can be derived from the new  effective Lagrangian
\be\label{lagefff}
 {\cal L}_{\rm eff} = \frac{1}{f}\bigg(
- \dot \varphi^2+\dot \phi^2\bigg)
-fV_{\rm eff}\, ,\qquad
V_{\rm eff}=J^2e^{-\mu\phi}+  2m^2 e^{\lambda \phi + 2\alpha \varphi}\ .
\ee
Let us now introduce new variables $\sigma_+=\varphi+\phi$, $\sigma_-=\varphi-\phi$. 
If a superpotential $\WW (\sigma_+,\sigma_-)$ exists such that
\be \label{supereq}
4\partial_+ \WW \partial_- \WW= V_{\rm eff} \ ,
\ee
then, the first-order system
\be\label{primerorden}
\dot\sigma_+ =2f\partial_- \WW\ ,\qquad \dot\sigma_-= 2f\partial_+ \WW\ ,
\ee
automatically solves the second-order equations of motion derived from 
${\cal L}_{\rm eff}$.
Our  strategy is to choose particular parameters
so that $V_{\rm eff} $ factorizes as $h_+(\sigma_+)h_-(\sigma_-)$.
In this case (\ref{supereq}) can be readily integrated.
The factorization occurs in two cases:
\bea
&& a)\ \mu=2\alpha-\lambda\ \ \longrightarrow \ \  V_{\rm eff}=e^{\alpha(1-\hat\lambda) \sigma_-} \left( J^2 e^{-\alpha (1-\hat\lambda)\sigma_+}+2m^2 e^{\alpha (1+\hat\lambda)\sigma_+}\right)
\nonumber\\
\nonumber\\
&& b)\ \mu=-2\alpha-\lambda \ \longrightarrow \ \  V_{\rm eff}
=e^{\alpha(1+\hat\lambda) \sigma_+} \left( J^2 e^{-\alpha (1+\hat\lambda)\sigma_-}+2m^2 e^{\alpha (1-\hat\lambda)\sigma_-}\right)
\nonumber
\eea
where $\hat\lambda\equiv \lambda/(2\alpha)$.
The superpotentials are 
\be
\WW^{(a)}=
\frac{2m^2}{\alpha(1+\hat\lambda)}e^{\alpha (1+\hat\lambda)\sigma_+}-\frac{J^2}{\alpha(1-\hat\lambda)}e^{-\alpha (1-\hat\lambda)\sigma_+}+\frac{1}{4\alpha(1-\hat\lambda)}e^{\alpha(1-\hat\lambda) \sigma_-}
\ee
\be
\WW^{(b)}=
\frac{2m^2}{\alpha(1-\hat\lambda)}e^{\alpha (1-\hat\lambda)\sigma_-}-\frac{J^2}{\alpha(1+\hat\lambda)}e^{-\alpha (1+\hat\lambda)\sigma_-}+\frac{1}{4\alpha(1+\hat\lambda)}e^{\alpha(1+\hat\lambda) \sigma_+}
\ee
It is important to note that this is an example where the superpotential is {\it not} of the form
${\cal W}=e^{\alpha\varphi } W(\phi_i)$, considered in previous sections,  but it has a more complicated dependence on $\varphi$. 


It is now easy to directly integrate the first-order equations (\ref{primerorden}). We choose  $f =e^{  -\alpha \varphi}$.
The results are as follows:

\subsubsection*{Case a)} 

In this case we define $\sigma_+=\tau$, so that $\sigma_-=2\varphi-\tau$, $\phi=\tau-\varphi $. Integrating the first-order equations we get

\be
\label{primaeq}
e^{2\alpha  (1-\hat\lambda ) \varphi(\tau)}=\frac{8m^2 (1-\hat\lambda )}{  
   1+\hat\lambda } \ e^{2 \alpha \tau }+b\ e^{\alpha  (1-\hat\lambda ) \tau }-4 J^2 \ ,
\ee
\be
 dt=2  d\tau\, e^{\alpha (1-\hat\lambda)\tau } e^{ \alpha(2 \hat\lambda -1) \varphi(\tau)}\ ,\qquad a=e^{\beta \varphi}\ ,\qquad H=\beta \partial_\tau\varphi\, \frac{d\tau}{dt}\ .
\ee

For $\hat \lambda<1$, this describes an expanding universe  starting at some finite $\tau=\tau_0$ where $\varphi\to -\infty$ and approaching the attractor solution $a\sim t^{1/\lambda^2}$ at $t\to\infty $.
This attractor is the asymptotic behavior of a large class of solutions when $\hat\lambda<1$.
The expansion is  accelerated at late times.

For $\hat\lambda>1$, the solution exists provided $b>0$. The solution starts at $\tau=-\infty$ where also $t\to -\infty$ and terminates at a finite $\tau_1$, where $\varphi\to\infty$ and $t\to\infty$.
The cosmology describes an expanding universe also in this case. There can be a transient period of acceleration, depending on the parameter values, but the cosmology undergoes a decelerated expansion at late times.

\subsubsection*{Case b)}

Now we define $\sigma_-=\tau$, so that $\sigma_+=2\varphi-\tau$, $\phi=\varphi-\tau $. We get

\be
\label{unamas}
e^{2\alpha  (1+\hat\lambda ) \varphi(\tau)}=
\frac{8m^2 (1+\hat\lambda )}{  
   1-\hat\lambda } \ e^{2 \alpha \tau }+b\ e^{\alpha  (1+\hat\lambda ) \tau }-4 J^2 \ ,
\ee
\be
 dt=2  d\tau\, e^{\alpha (1+\hat\lambda)\tau } e^{- \alpha(2 \hat\lambda +1) \varphi(\tau)}\ ,\qquad a=e^{\beta \varphi}\ ,\qquad H=\beta \partial_\tau\varphi\, \frac{d\tau}{dt}\ .
\label{ultima}
\ee
For $\hat \lambda<1$, the cosmology is qualitatively similar to the case a), approaching the attractor solution $a\sim t^{1/\lambda^2}$ at late times. For $\hat \lambda>1$, existence of the solution requires $b>0$. In this case the cosmology starts at some finite $\tau$, with the asymptotic late-time behavior $\varphi\sim \frac12 \tau \sim \frac{1}{\alpha} \log t$, so that
$a\sim t^{1/3}$.

\medskip

The case $\hat\lambda =1$ is special and needs to be discussed separately (the case  $\hat\lambda =-1$ is equivalent by a field redefinition $\phi\to -\phi$). 
In the case a), when $\hat \lambda =1$, one has $\mu=0$.
Then the field $\chi$ has no dilaton coupling in the kinetic
term. Since our interest is to investigate the role of the dilaton coupling, here we will focus on case b).\footnote{ In particular, the  case a) with $\hat\lambda=1$ and $\dot \chi=0$ reduces to the scalar field with exponential potential
whose general solution is given in \cite{Russo:2004ym}.}
The superpotential now becomes
\be
\hat\lambda=1:\quad \WW^{(b)}=
2m^2\sigma_--\frac{J^2}{2\alpha}e^{-2\alpha \sigma_-}+\frac{1}{8\alpha}e^{2\alpha \sigma_+} \ .
\ee
Defining again $\sigma_-=\tau$,  $\sigma_+=2\varphi-\tau$, $\phi=\varphi-\tau $, one obtains the solution

\be
\label{unitamas}
e^{4\alpha   \varphi(\tau)}=
16\alpha m^2 \tau  e^{2 \alpha \tau }+b\ e^{2\alpha  \tau }-4 J^2 \ ,
\ee
\be
 dt=2  d\tau\, e^{2\alpha\tau } e^{- 3\alpha \varphi(\tau)}\ ,\qquad a=e^{\beta \varphi}\ ,\qquad H=\beta \partial_\tau\varphi\, \frac{d\tau}{dt}\ .
\label{ultimita}
\ee
This again describes an expanding cosmology starting at some $\tau=\tau_0$ and approaching the attractor  $a\sim t^{1/3}$ at late times (which is the limiting value of $a\sim t^{1/\lambda^2}$ for $\hat \lambda \to 1$, {\it i.e.} $\lambda\to \sqrt{3}$).

A variant of this model where $\phi $ or $\chi $ are phantom fields has been investigated in \cite{Paliathanasis:2020sfe}, where the general properties of the resulting cosmologies were exhibited, by studying trajectories and fixed points. In phantom models the equation of state parameter $w$ can cross the $w=-1$ barrier.
Using the present formalism, we can now also provide  solutions in closed form in the case when  $\chi$ is a phantom field. We take the same Lagrangian \eqref{teorla}, but with the opposite sign in the kinetic term for $\chi $.
This has the effect of reversing the sign of $J^2$ in the effective potential \eqref{lagefff}. Thus the above results \eqref{primaeq}-\eqref{ultima} apply with the formal change  $J^2\to -J^2$. 

For $\hat \lambda<1$, the late time behavior of the resulting cosmologies is the same as in the previous cases, approaching the attractor solution $a\sim t^{1/\lambda^2}$ at late times.
However, now the scale factor approaches a constant value at $\tau\to-\infty$ and the expansion begins with $\dot a=0$. 
In case a), for $b>0$ and $\hat \lambda>1$, the early and late time behavior are the same as in \eqref{primaeq}. 
However, in case b), for $b>0$ the scale factor  approaches a constant value at $\tau\to-\infty$.   Solutions with $b<0$ can now exist for $\hat\lambda>1$ provided $J^2$ is above a critical value. This is determined by the condition that the RHS of \eqref{unamas} (with $J^2\to-J^2$) is positive, which for $b<0$ and $\hat\lambda>1$ is possible in a finite $\tau $ interval.


\medskip

In conclusion, one can describe interesting axion-dilaton cosmologies in terms of analytic formulas. These new exact solutions 
depend on two integration constants $J$, $b$ and on the parameters of the potential $m$ and $\lambda$. By varying these parameters one can describe many different cosmologies with features of interest, such as late time or transient acceleration. The solution also exhibits the transfer of kinetic energies between the axion $\chi $ and $\phi $
due to their coupling, which takes place along the time evolution.

\section{Summary and Discussion}

The prevailing $\Lambda$CDM model of cosmology incorporates a small positive cosmological constant. This constant leads to the emergence of a de Sitter geometry during the late stages of cosmic evolution. The interpretation of this model as an effective theory derived from the compactification of string/M-theory requires the existence of a compactification to  a four-dimensional de Sitter universe. However, thus far, no definitive ``de Sitter compactification" has been identified. Nevertheless, the observed evidence for accelerated cosmic expansion can still be attributed to a string compactification that yields a cosmological spacetime resembling de Sitter for a sufficiently extended period.

Numerous attempts have been made over the years to generate inflation through various mechanisms. However, an attractive and  natural realization of inflation is still based on the coupling of Einstein gravity to scalar fields with  self-interactions driven by a potential. 
This viewpoint not only offers a way  to analytically formulating the inflationary scenario but also possesses conceptual appeal by establishing a direct connection between Cosmology and the Fundamental Interactions. Scalar fields, which are widely present and abundant in Supergravity across different dimensions, encode properties of specific compactifications and thereby convey insights from Fundamental Physics.

Finding exact result solutions in interacting theories with multiple scalar fields is 
in general extremely complicated. In this paper we revisited the method of the Superpotential, finding
a number of new interesting examples.
Sections 3 and section 4 study models with canonical kinetic terms for the scalar fields ({\it viz.} with flat target metric). In particular, we found  novel exact solutions that interpolate between distinct de Sitter phases of the Universe. 
Specifically, in the two-scalar field model detailed in Section 4.3, these solutions describe trajectories within a two-dimensional space, traversing a remarkably diverse landscape of peaks and valleys that gradually changes as the coupling constants are varied. The dynamics is characterized in terms of the various fixed points and an interesting intuitive picture emerges. Importantly, the derived exact solution provides explicit expressions detailing the time-dependence of the scale factor.

We have also described the construction of exact domain-wall solutions from the cosmological solutions. Such solutions interpolate between AdS metrics and are potentially interesting for holographic applications,
as they describe a renormalization group flow between
 two different CFT's.

In Section 5, we have examined an illustrative example that shows the construction of solutions within axion-dilaton models. The target metric governing the kinetic terms now describes a hyperbolic 2-space parametrized by $(\phi,\chi)$. This construction exhibits an alternative approach to obtaining exact solutions. By integrating out the axion field, an effective potential for a single scalar field emerges, albeit with a non-factorized dependence on the metric component $e^{\alpha\varphi}$. Nevertheless, we can still solve the equation for the superpotential by observing that, for specific values of the coupling $\mu$, it assumes a remarkably simple form in terms of newly introduced scalar field variables  $\sigma_+$ and $\sigma_-$. Consequently, three novel solutions for the axion-dilaton model arise, given in \eqref{primaeq}, \eqref{unamas}. \eqref{unitamas}. These solutions describe expanding universes that, in a certain parameter range, exhibit transient periods of acceleration.
The derived solutions provide valuable insights into the underlying mechanisms responsible for acceleration, offering analytical control over the process. There are interesting potential applications to other aspects of cosmology. In particular, 
it would be interesting to perform a comprehensive  analysis aimed at determining  the  parameter range required for  the transient periods of acceleration to approximate de Sitter spacetime over an extended duration, in order to produce the required number of e-foldings needed to solve the horizon problem.
Furthermore, investigating potential connections between the observed special values of the coupling parameter $\mu$ and specific supergravity compactifications would be of great interest.

\subsection*{Acknowledgments}
We would like to thank J. Garriga and P. Townsend for useful  comments.
We acknowledge financial support from 
 grants 2021-SGR-249 (Generalitat de Catalunya), and  MINECO
 PID2019-105614GB-C21.

\end{document}